\documentclass[preprint,showpacs,preprintnumbers,amsmath,amssymb]{revtex4}
\usepackage{graphicx}
\usepackage{dcolumn}
\usepackage{bm}
\usepackage{mathrsfs}

\begin{document}

\title{Fermion Absorption Cross Section and Topology of Spherically Symmetric Black Holes}

\author{Yu-Xiao Liu}
\thanks{Corresponding author}%
\email{liuyx@lzu.edu.cn}%
\author{Li Zhao}
\author{Zhen-Bin Cao}
\author{Yi-Shi Duan}
\affiliation{Institute of Theoretical Physics, Lanzhou University,
Lanzhou 730000, P. R. China}

\begin{abstract}
In 1997, Liberati and Pollifrone in Phys. Rev. D56 (1997) 6458
(hep-th/9708014) achieved a new formulation of the
Bekenstein-Hawking formula, where the entropy and the Euler
characteristic are related by $S = \chi {\cal A}/8$. In this work
we present a relation between the low-energy absorption cross
section for minimally coupled fermions and the Euler
characteristic of (3+1)-dimensional spherically symmetric black
holes, i.e. $\sigma =\chi~g_h^{-1}{\cal A}$. Based on the
relation, using the Gauss--Bonnet--Chern theorem and the
$\phi$-mapping method, an absorption cross section density is
introduced to describe the topology of the absorption cross
section. It is shown that the absorption cross section and its
density are determined by the singularities of the timelike
Killing vector field of the spacetime and these singularities
carry the topological numbers, Hopf indices and Brouwer degrees,
naturally.
\end{abstract}


\maketitle

\section{Introduction}\label{Sec1}

Interest in the absorption of quantum waves by black holes was
reignited in the 1970s, following Hawking's discovery that black
holes can emit, as well as scatter and absorb, radiation
\cite{haw75}. More recently, absorption cross sections have been
of interest in the context of higher-dimensional string theories.

In a series of papers \cite{san77,san78a,san78} Sanchez considered
the scattering and absorption of massless scalar particles by an
uncharged, spherically-symmetric black hole. Unruh \cite{unr76}
studied the absorption of massive spin-half particles by piecing
together analytic solutions to the Dirac equation across three
regions. He showed that, in the low-energy limit, the scattering
cross section for the fermion is exactly $1/8$ of that for the
scalar particle. Doran {\em et al.} \cite{DoranPRD200571} studied
the absorption of massive spin-half particles by a small
Schwarzschild black hole by numerically solving the
single-particle Dirac equation in Painlev\'e--Gullstrand
coordinates.

In Ref. \cite{DasPRL199778}, Das {\em et al.} computed the
low-energy absorption cross section for minimally coupled massless
scalars and spin-half particles, into a general spherically
symmetric black hole in arbitrary dimensions. It is interesting to
note for extremal black holes, the absorption cross section for
minimally coupled fermion vanishes in the limit of extremality. In
2005, Jung, Kim and Park \cite{JungPLB2005614} computed the ratio
of the low-energy absorption cross section for Dirac fermion to
that for minimally coupled scalar when the spacetimes are various
types of the higher-dimensional spherically symmetric black holes.
They found that the low-energy absorption cross section for the
Dirac fermion always goes to zero in the extremal limit regardless
of the detailed geometry of the spacetime. So, there should be
some relation between the low-energy absorption cross section for
minimally coupled fermions and the topology of spherically
symmetric black holes. In this paper we will consider the relation
between them.

This paper is organized as follows. In Section \ref{Sec2}, using
the Gauss--Bonnet--Chern (GBC) theorem, the Euler characteristic
and the low-energy absorption cross section for Dirac fermions of
both the extremal and nonextremal spherically symmetric black
holes are briefly reviewed. Some useful notations are also
prepared. In Section \ref{Sec3}, from the opinion of the
decomposition of spin connection, the density of the absorption
cross section of spherically symmetric black holes is proposed in
terms of the smooth unit tangent vector field. Then, in Section
\ref{Sec4}, using the $\phi$-mapping method
\cite{DuanMengJMP1993}, the topology of the absorption cross
section is investigated at the singularities of the unit vector
field. In Section \ref{Sec5}, we compute the value of the Hopf
index for several example spacetimes. The conclusion of this paper
is given in Section \ref{SecConclusions}.

\section{The Euler characteristic and the absorption cross section
for Dirac Fermion of spherically symmetric black holes}
\label{Sec2}

Let us consider the general spherically symmetric metric in
$(3+1)$-dimensional spacetime of the form
\begin{equation}
 ds^2 = f(\hat{r}) dt^2 - g(\hat{r}) \left(d\hat{r}^2 + \hat{r}^2 d\Omega^2
\right),\label{general1}
\end{equation}
where $d\Omega^2 =d\theta^2+\sin^{2}{\theta}d\varphi^2$ is the
metric on the unit 2-sphere. We also assume
\begin{equation}
 \lim_{\hat{r} \rightarrow \infty} f(\hat{r})
 = \lim_{\hat{r} \rightarrow \infty} g(\hat{r}) = 1
\end{equation}
for ensuring that the metric is asymptotically flat. Ref.
\cite{DasPRL199778} has shown that the low-energy absorption cross
section for minimally coupled fermions is given by the area
measured in the flat spatial metric conformally related to the
true metric in the form
\begin{equation}
\sigma = 2 g_h^{-1}{\cal A},
\label{blfermion}
\end{equation}
where ${\cal A} \equiv 4\pi R_h^2 $ is the horizon area, $R_h =
\hat{r}_h \sqrt{g_h}$, $g_h=g(\hat{r}_h)$, and $\hat{r}_h$ is a
horizon radius usually determined by the largest solution of
$f(\hat{r})=0$, the factor $2$ comes from the number of spinors.

As a fundamental but very important case of (\ref{general1}), the
Reissner--Nordstr\"{o}m (RN) black hole has the metric form
\begin{equation}
 ds_{RN}^2=F(r)dt^{2} -F(r)^{-1}dr^{2}
     -r^{2}d\Omega^{2}.\label{metricRN}
\end{equation}
with
\begin{equation}
F(r)=\left(1-\frac{2M}{r}+\frac{Q^2}{r^2}\right).
\end{equation}
There are two horizons, the event horizon $r_+$ and Cauchy horizon
$r_{-}$,
\begin{equation}\label{horizonRN}
r_{\pm}= M \pm \sqrt{M^2 -Q^2}.
\end{equation}
The extremal RN black hole corresponds to the case $r_{+} = r_{-}$
or, equivalently, $M=Q$. The metric (\ref{metricRN}) can also be
written in the form \cite{JungPLB2005614}
\begin{equation}
 ds_{RN}^2 = f_{RN}(\hat{r}) dt^2 - g_{RN}(\hat{r}) \left(d\hat{r}^2
             + \hat{r}^2 d\Omega^2 \right) \label{metricRN2}
\end{equation}
with
\begin{eqnarray}
 f_{RN}(\hat{r})
  &=& \left[ \frac{  \hat{r}^2  - {\cal C}^2 }
                  { \left( \hat{r}  + {\cal C} \right)^2
                    + \hat{r} r_{-} }
      \right]^2,   \\  \nonumber
 g_{RN}(\hat{r}) &=&
 \left[ \frac{( \hat{r} + {\cal C})^2}{\hat{r}^2}
        + \frac{r_{-}}{\hat{r}}
 \right]^2,
\end{eqnarray}
where ${\cal C} = (r_{+}-r_{-})/4$, and the relation between $r$
and $\hat{r}$ is
\begin{equation}
 r = \frac{(\hat{r}+{\cal C})^2}{\hat{r}} + r_{-}.
\end{equation}

Using the GBC theorem and the boundary conditions, it is shown
that the Euler characteristic
\begin{equation}\label{nonextremal}
\chi= 2
\end{equation}
for the nonextremal RN black hole
\cite{Zaslavskii1996,MaldacenaPRL1996,Zaslavskii1997} and
\begin{equation}\label{extremal}
\chi= 0
\end{equation}
for the extremal RN black hole \cite{Hawking1995,Teitelboim1995}.
In the later sections, we will prove that the above two formulas
are held for general spherically symmetric black holes as well as
for the Kerr and Kerr--Newman black holes.

It is interesting to note that the low-energy absorption cross
section $\sigma$ for the Dirac fermion always vanishes in the
extremal limit regardless of the detailed geometry of the
spacetime \cite{DasPRL199778,JungPLB2005614}. So we introduce the
relation between the absorption cross section and the topology of
spherically symmetric black holes
\begin{equation}\label{relation}
\sigma= g_h^{-1}{\cal A}~\chi.
\end{equation}
Thus, the topologies (\ref{nonextremal}) and (\ref{extremal}) of
spherically symmetric black holes lead directly to the nonzero and
zero results of the absorption cross section by taking account of
the relation (\ref{relation}).

\section{The absorption cross section density in spherically symmetric black holes}\label{Sec3}

The topologies (\ref{nonextremal}) and (\ref{extremal}) give the
global structure of spherically symmetric black holes. In this
section, in order to show the relation between the absorption
cross section and the topology of spherically symmetric black
holes, let us consider the Euler characteristic in detail and
introduce the absorption cross section density of spherically
symmetric black holes.

For a closed $N$(even)-dimensional Riemannian  manifold $M^{N}$,
the Euler characteristic $\chi(M^{N})$ can be expressed as the
volume integral of the GBC differential $N$-form $\Lambda$:
\begin{equation}
\chi(M^{N})=\int_{M^{N}}\Lambda,\label{GBC}
\end{equation}
\begin{equation}
 \Lambda={\frac{(-1)^{N/2-1}}{2^N \pi ^{N/2} \left(\frac{N}{2}\right)!}}
         \varepsilon_{A_1A_2\cdots {A_{N-1}A_N}}F^{{A}_1{A}_2}
         \wedge \cdots \wedge {F^{{A}_{N-1}{A}_N}},
\end{equation}
in which $F^{AB}$ is the curvature tensor of $SO(N)$ principal
bundle of the Riemannian manifold $M$, i.e., the $SO(N)$ gauge
field  2-form
\begin{equation}\label{FAB}
F^{AB}=d\omega ^{AB}-\omega ^{AC}\wedge \omega ^{CB},
\end{equation}
where $\omega ^{AB}$ is the spin connection $1$-form. Chern has
shown that the GBC $N$-form $\Lambda$ on $M^{N}$ can be pulled
back to $S(M^{N})$ as the exterior derivative of a differential
$(N-1)$-form $\Omega$:
\begin{equation}
\pi^{*}\Lambda=d\Omega,
\end{equation}
where $\pi^{*}$ denotes the pullback of the projection
$\pi:S(M^{N})\rightarrow M^{N}$. Using a recursion method, Chern
has proved that the $(N-1)$-form on $S(M^{N})$ can be written as
\begin{equation}
 \Omega = \frac{1}{\left(2\pi \right)^{N/2}}
          \sum\limits_{k = 0}^{N/2 - 1} {( - 1)^k {{2^{ - k} }
          \over {(N - 2k - 1)!!k!}}\Theta _k },
\end{equation}
which is called the Chern form with
\begin{equation}\label{theta1}
\begin{array}{ll}
 \Theta _k =& \varepsilon _{A_1 A_2 \cdots A_{N - 2k}
              A_{N - 2k +1} A_{N - 2k + 2} \cdots A_{N - 1} A_N }
              n^{A_1 } \theta ^{A_2}\wedge \cdots \vspace{.4cm} \\
            & \wedge \theta ^{A_{N - 2k} }
              \wedge F^{A_{N - 2k + 1} A_{N - 2k + 2} }
              \wedge \cdots \wedge F^{A_{N - 1} A_N },
\end{array}
\end{equation}and
\begin{equation}
\theta ^A \equiv Dn^A = dn^A - \omega ^{AB} n^B,
\end{equation}
in which $n^A$ is the section of the sphere bundle $S^{N-1}(M)$
\begin{equation}
n:\partial M \to S^{N - 1}(M).
\end{equation}
It is noted that $\pi^{*}$ maps the cohomology of $M^{N}$ into
that of $S(M^{N})$, where $n^{*}$ performs the inverse operation.
Thus $n^{*}\pi^{*}$ amounts to the identity and the Euler
characteristic $\chi(M^{N})$ in Eq. (\ref{GBC}) can be written as
\begin{equation}
 \chi(M^{N})=\int_{M^{N}}\Lambda
            =\int_{M^{N}}n^{*}\pi^{*}\Lambda
            =\int_{M^{N}}n^{*}d\Omega.
\end{equation}
In the opinion of the decomposition of gauge potential, Duan {\em
et al.} (\cite{DuanMengJMP1993,DuanLeeHPA1995,YangDuanMPLA1998})
showed that the $(N-1)$-form can be formulated in terms of the
unit tangent vector field $n^{a}(x)$ cleanly as
\begin{equation}
 \Omega=\frac{\Gamma^{N/2}}{(N-1)!(2\pi)^{N/2}}\epsilon_{a_{1}a_{2}
        {\cdots} a_{N}} n^{a_{1}}dn^{a_{2}}\wedge \cdots \wedge dn^{a_{N}}.
\end{equation}
Therefore the Euler characteristic $\chi(M^{N})$ is
\begin{eqnarray}
 \chi(M^{N})&=&\frac{1}{(N-1)!A(S^{N-1})}\int_{M^{N}}\epsilon^{\mu_{1}\mu_{2}
            {\cdots} \mu_{N}} \epsilon_{a_{1}a_{2}
            {\cdots} a_{N}} \nonumber \\
            &&\times\partial_{\mu_{1}}n^{a_{1}}\partial_{\mu_{2}}n^{a_{2}}
            {\cdots} \partial_{\mu_{N}}n^{a_{N}} d^{N}x,\label{EulerCharacteristic}
\end{eqnarray}
where $A(S^{N-1})$ is the surface area of $(N-1)$-dimensional unit
sphere $S^{N-1}$,
\begin{equation}
A(S^{N-1})=\frac{(2\pi)^{N/2}}{\Gamma^{N/2}}.
\end{equation}

For (3+1)-dimensional spherically symmetric black holes, Eq.
(\ref{EulerCharacteristic}) becomes
\begin{equation}\label{EulerCharacteristic RN}
 \chi=\frac{1}{12\pi^2}\int_{M}\epsilon^{\mu\nu\lambda\rho}
         \epsilon_{abcd}\partial_{\mu}n^{a}\partial_{\nu}n^{b}
         \partial_{\lambda}n^{c}\partial_{\rho}n^{d} d^{4}x,
\end{equation}
and $n^{a}(x)$ coincides with the timelike Killing vector field of
the spherically symmetric spacetime. According to the relation
(\ref{relation}), we introduce the absorption cross section
density $\rho$ of spherically symmetric black holes
\begin{equation}\label{density}
 \rho=\frac{\cal A}{g_h}\cdot \frac{1}{12\pi^2}
      \epsilon^{\mu\nu\lambda\rho}
      \epsilon_{abcd}\partial_{\mu}n^{a}\partial_{\nu}n^{b}
      \partial_{\lambda}n^{c}\partial_{\rho}n^{d}.
\end{equation}
Then the absorption cross section $\sigma$ of spherically
symmetric black holes reads as
\begin{equation}\label{CrossSection2}
\sigma=\int_{M}\rho ~d^{4} x.
\end{equation}
In the following, we will consider the topology of the absorption
cross section of spherically symmetric black holes through the
absorption cross section density $\rho$ and the so-called
$\phi$-mapping method.

\section{The topology of the absorption cross section in spherically symmetric black holes}\label{Sec4}

The unit tangent vector $n^{a}$ over $M$ satisfies
\begin{equation}
n^{a}n^{a}=1, ~~~~~ a=1,2 \cdots 4,
\end{equation}
and can, in general, be further expressed as
\begin{equation}\label{unit vector}
n^{a}=\frac{\phi^{a}}{\|\phi\|},\;\|\phi\|=\sqrt{\phi^{a}\phi^{a}},
\end{equation}
where $\phi^{a}=e^{a}_{\mu}\phi^\mu$, $e^a_\mu$ and $\phi^{\mu}$
are the vierbein and the timelike Killing vector field of
spherically symmetric black holes, respectively. Substituting
(\ref{unit vector}) into (\ref{density}) and considering that
\begin{equation}
 \partial_\mu n^a=\frac{\partial_\mu \phi ^a}{\left\| \phi \right\|}
                  +\phi^a\partial_\mu \frac{1}{\left\|\phi\right\|},
\end{equation}
we have
\begin{equation}
 \rho=-\frac{\cal A}{24\pi^{2}g_{h}}
      \epsilon^{\mu\nu\lambda\rho}
      \epsilon_{abcd}\partial_{\mu}\phi^{k}\partial_{\nu}\phi^{b}
      \partial_{\lambda}\phi^{c}\partial_{\rho}\phi^{d}
      \frac{\partial}{\partial{\phi^{k}}}
      \frac{\partial}{\partial{\phi^{a}}}
      \frac{1}{\|\phi\|^2}.
\end{equation}
If we define the Jacobian $J(\phi/x)$ as
\begin{equation}\label{1-firstjac}
 \epsilon^{kbcd}J(\phi/x) = \epsilon^{\mu\nu\lambda\rho}
        \partial_{\mu}\phi^{k}\partial_{\nu}\phi^{b}
        \partial_{\lambda}\phi^{c}\partial_{\rho}\phi^{d}
\end{equation}
and notice $\epsilon_{abcd}\epsilon^{kbcd}=3!\delta^k_a$, we
obtain
\begin{equation}
 \rho=-\frac{\cal A}{4\pi^{2}g_{h}}
      \left(\frac{\partial^{2}}{\partial{\phi^{a}} \partial{\phi^{a}}}
      \frac{1}{\|\phi\|^2} \right)
       J(\frac{\phi}{x}).
\end{equation}
Via the general Green function relation in $\phi$-space
\cite{Gelfand1958}
\begin{equation}
 \frac{\partial^2}{\partial{\phi^{a}} \partial{\phi^{a}}}
 \frac{1}{\|\phi\|^2} =-4\pi^2 \delta(\phi),
\end{equation}
we do obtain the $\delta$-function like absorption cross section
density $\rho$ of spherically symmetric black holes
\begin{equation}\label{density_delta}
\rho=\frac{\cal A}{g_h}~\delta(\phi) J(\frac{\phi}{x}).
\end{equation}
From the above formula one see that the absorption cross section
density does not vanish only at the zero points of the vector
field ${\phi}(x)$. This result shows that the topology of the
absorption cross section is determined by the zeros of $\phi(x)$,
i.e. the singularities of $n(x)$ which coincides with the timelike
Killing vector field of the spherically symmetric spacetime.
Therefore, it is essential to investigate the solutions of
${\phi}(x)=0.$

Suppose that the vector field ${\phi}(x)$ has l zero points
which are denoted as $\vec{z}_i$ $(i=1,\cdots,l).$ According to
the implicit function theorem \cite{Goursat1904}, when the zero
points $\vec{z}_i$ are the regular points of ${\phi}(x),$ i.e.
when the Jacobian determinant
\begin{equation}
\left. J\left( \frac \phi x\right) \right| _{\vec{z}_i} =\left.
\frac{\partial (\phi^1,\phi^2,\phi^3,\phi^4)}{\partial
(x^1,x^2,x^3,x^4)}\right| _{\vec{z}_i}\neq 0,
\end{equation}
where $J\left( \phi /x\right) =J^0\left( \phi /x\right) $ is the
usual Jacobian determinant, there exists one and only one
continuous solution of $\phi ^a(x)=0$. The solution can be
expressed as $\vec{z}_i=\vec{z}_i(t)$, which is the trajectory of
the {\em i}th zero point of ${\phi}(x)$. According to the
$\delta$-function theory \cite{Schouten1951} and the
$\phi$-mapping theory, we know that $\delta(\phi)$ can be expanded
by these zeros as
\begin{equation} \label{delta expande}
\delta(\phi)=\sum_{i=1}^l\frac{\beta_i} {\left| J(\phi/x)\right|
_{\vec{z}_i}}\delta(\vec{x}- \vec{z}_i),
\end{equation}
where the positive integer $\beta_i$ is called the Hopf index of
the $\phi$-mapping at $\vec{z}_i$ and it means that, when the
point $\vec{x}$ covers the neighborhood of $\vec{z}_i$ once, the
function $\phi(x)$ covers the corresponding region $\beta_i$
times, which is a topological number of first Chern class and
relates to the generalized winding number of the $\phi$-mapping.
Substituting (\ref{delta expande}) into (\ref{density_delta}), the
absorption cross section density $\rho$ is formulated by
\begin{equation}
\rho=\frac{\cal A}{g_h}~\sum_{i=1}^l {\beta_i}{\eta_i}
\delta(\vec{x}-\vec{z}_i),\label{density_beta_eta}
\end{equation}
where $\eta_i$ is called the Brouwer degree of the $\phi$-mapping
at $\vec{z}_i$ \cite{DuanMengJMP1993,Hopf1929} and
\begin{equation}
\eta_i= \texttt{sign} J(\phi/x)|_{\vec{z}_i}=\pm 1
\end{equation}
according to the clockwise or anti-clockwise rotation of $\phi(x)$
when $\vec{x}$ circles $\vec{z}_i$ clockwise. So, from
(\ref{CrossSection2}), the absorption cross section $\sigma$ of
spherically symmetric black holes is given by the sum of these
Hopf indices and Brouwer degrees of the Killing vector field at
its singularities, i.e.
\begin{equation}
\sigma=\frac{\cal A}{g_h}~\sum_{i=1}^l
{\beta_i}{\eta_i},\label{CrossSection_beta_eta}
\end{equation}
which is the direct result of the combination of the relation
(\ref{relation}) and the Hopf index theorem.

From (\ref{density_beta_eta}) and (\ref{CrossSection_beta_eta}) we
see that the absorption cross section density is determined by the
singularities of the timelike Killing vector field and the
absorption cross section is characterized by the topological
numbers, Hopf indices and Brouwer degrees of $\phi$-mapping, at
these singularities.

\section{The value of the Hopf index for several example spacetimes}\label{Sec5}
In this section, we compute the value of the Hopf index for
several example black holes and for a general static spherically
symmetric metric.

\subsection{The Schwarzschild black hole}
The Schwarzschild black hole has the metric
\begin{equation}
 ds^2=\left(1-\frac{2M}{r}\right)dt^{2}
     -\left(1-\frac{2M}{r}\right)^{-1}dr^{2}
     -r^{2}d^{2}\Omega.\label{metricSchwarzschild}
\end{equation}
with the horizon located at $r=2M$. Using the GBC theorem, it has
been shown (Liberati and Pollifrone, 1997) that the Euler
characteristic of the Schwarzschild black hole is
\cite{LiberatiPRD1997}
\begin{equation}
\chi=2,\label{chiSchwarzschild}
\end{equation}
which leads directly to the Bekenstein--Hawking entropy $S = {\cal
A}/4$ by taking account of the relation $S = \chi {\cal A}/8$. The
topology (\ref{chiSchwarzschild}) gives the global property of the
Schwarzschild black hole. In the following, let us compute the
value of the Hopf index for the Schwarzschild black hole.

By using the Killing equation
\begin{equation}
 (\partial_\lambda g_{\mu\nu})\phi^\lambda
 + g_{\lambda\mu} \partial_\nu \phi^\lambda
 + g_{\lambda\nu} \partial_\mu \phi^\lambda=0,\label{KillingEq}
\end{equation}
and the static and spherically symmetric properties of the
Schwarzschild black hole that imply
\begin{equation}
 \partial_0 \phi^\mu = \partial_3 \phi^\mu, ~~~ \mu=0,1,2,3,
 \label{KillingEqCondiction}
\end{equation}
one can solve the Killing vector field $\phi^\mu(x)$ and, by
transforming into the local orthonormal vierbein index ``a":
$\phi^a=e^a_{\mu}\phi^\mu(x)$, one obtains
\begin{eqnarray}
 \phi^0=\sqrt{1-\frac{2M}{r}},~~~
 \phi^1=\phi^2=0,~~~
 \phi^3=r\sin\theta.
 \label{EulerCharacteristic}
\end{eqnarray}
It is easy to see that for the Schwarzschild black hole the
Killing vector field $\phi^a$ has two zeros located at
\begin{equation}
 (r=2M,\theta=0), ~~~ (r=2M,\theta=\pi).
 \label{zerosSchwarzschild}
\end{equation}
One can see the distribution of Killing vector field in the
figures in Ref. \cite{yangCTP2005}. For the zero ($2M,0$), the
Killing vector field $\phi^a$ rotates from 0 to $\pi$
anti-clockwise when the spacetime point ($r,\theta$) circles the
zero in the same way. So we have the Hopf index and the Brouwer
degree
\begin{equation}
 \beta_1=1,~~~ \eta_1=+1,
 \label{zerosSchwarzschild}
\end{equation}
at the zero ($2M,0$). Similarly, for the zero ($2M,\pi$), we get
\begin{equation}
 \beta_2=1,~~~ \eta_2=+1.
 \label{zerosSchwarzschild}
\end{equation}
Then, from the general results in Eqs. (\ref{density_beta_eta})
and (\ref{CrossSection_beta_eta}), we obtain the Euler
characteristic and the absorption cross section of the
Schwarzschild black hole
\begin{eqnarray}
 \chi &=& \sum_{i=1}^2 {\beta_i}{\eta_i}=2, \label{ourChiSchwarzschild} \\
 \sigma &=& \frac{\cal A}{g_h}~\sum_{i=1}^2 {\beta_i}{\eta_i}
         = 2\frac{\cal A}{g_h},
 \label{ourSigmaSchwarzschild}
\end{eqnarray}
and particularly the absorption cross section density
\begin{equation}
\rho=\frac{\cal A}{g_h}~\delta(r-2M)
\left[\delta(\theta)+\delta(\theta-\pi)\right].
\label{ourRhoSchwarzschild}
\end{equation}

\subsection{The RN black hole}

For the RN black hole the metric is given in Eq. (\ref{metricRN}).
By using the Killing equation (\ref{KillingEq}) and the condition
for the Killing vector field of the RN black hole
(\ref{KillingEqCondiction}), we can also solve the Killing vector
field $\phi^\mu(x)$. The result for $\phi^a=e^a_{\mu}\phi^\mu(x)$
is of the form
\begin{eqnarray}
 \phi^0=\sqrt{1-\frac{2M}{r}+\frac{Q^2}{r^2}},~~~
 \phi^1=\phi^2=0,~~~
 \phi^3=r\sin\theta.
 \label{phiaRN}
\end{eqnarray}
For the nonextremal RN black hole, it turns out that the Killing
vector field $\phi^a$ has four zeros located at
\begin{equation}
\begin{array}{c}
  (r=r_{+},\theta=0), ~~~ (r=r_{-},\theta=0), \\
  (r=r_{+},\theta=\pi), ~~~ (r=r_{-},\theta=\pi), \\
\end{array}
\label{zerosRN}
\end{equation}
where $r_{\pm}= M \pm \sqrt{M^2 -Q^2}$. For the zero ($r_{+},0$),
the Killing vector field $\phi^a$ rotates from 0 to $\pi$
anti-clockwise when the spacetime point ($r,\theta$) circles the
zero in the same way. So we have the Hopf index and the Brouwer
degree
\begin{equation}
 \beta_1=1,~~~ \eta_1=+1,
\end{equation}
at the zero ($r_{+},0$). While for the zero $(r_{-},0)$, when
($r,\theta$) circles the zero from $\pi$ to 0 anti-clockwise,
$\phi^a$ rotates from $\pi$ to 0 clockwise, which leads to the
Hopf index and the Brouwer degree
\begin{equation}
 \beta_2=1,~~~ \eta_2=-1,
\end{equation}
Similarly, one can get
\begin{equation}
 \beta_3=1,~~~ \eta_3=-1,
\end{equation}
for the zero ($r_{-},\pi$) and
\begin{equation}
 \beta_4=1,~~~ \eta_4=+1,
\end{equation}
for the zero ($r_{+},\pi$). Since the event horizon is the
boundary of the outer spacetime of the RN black hole, it is only
the zeros ($r_{+},0$) and ($r_{+},\pi$) that have contribution to
the Euler characteristic and the absorption cross section. Then,
the Euler characteristic, the absorption cross section and its
density of the RN black hole are
\begin{eqnarray}
 \chi &=& \sum_{i=1,4} {\beta_i}{\eta_i}=2, \label{ourChiRN1} \\
 \sigma &=& \frac{\cal A}{g_h}~\sum_{i=1,4} {\beta_i}{\eta_i}
         = 2\frac{\cal A}{g_h},
 \label{ourSigmaRN1} \\
 \rho&=&\frac{\cal A}{g_h}~\delta(r-r_{+})
\left[\delta(\theta)+\delta(\theta-\pi)\right]. \label{ourRhoRN1}
\end{eqnarray}
respectively.

For the extremal RN black hole, the Killing vector field $\phi^a$
has two zeros located at
\begin{equation}
\begin{array}{c}
  (r=M,\theta=0), ~~~ (r=M,\theta=\pi). \\
\end{array}
\label{zerosRN}
\end{equation}
In this case, when the spacetime point ($r,\theta$) circles these
zeros anti-clockwise, $\phi^a$ rotates anti-clockwise at the outer
spacetime of horizon but clockwise at the inner of horizon. So we
have the Hopf indices
\begin{equation}
 \beta_1=\beta_2=0,
\end{equation}
which gives the Euler characteristic, the absorption cross
section, and its density of the extremal RN black hole
\begin{equation}
 \chi=0,~~~ \sigma=0,~~~ \rho=0.
\end{equation}
This result is in agreement with the statements that the extremal
black hole has a unique internal state \cite{Hawking1995} and its
temperature is zero due to the vanishing of surface gravity on
horizon.

\subsection{General spherically symmetric black holes}
In above two subsection, we have consider two example of
spherically symmetric black holes. Now let us compute the value of
the Hopf index for general spherically symmetric black holes.

In this general case, the metric can be taken the form given in
Eq. (\ref{general1}). By using the Killing equation
(\ref{KillingEq}) and the condition for the Killing vector field
(\ref{KillingEqCondiction}), we can also get the general form of
the Killing vector field $\phi^a$
\begin{eqnarray}
 \phi^0=\sqrt{F(r)},~~~
 \phi^1=\phi^2=0,~~~
 \phi^3=r\sin\theta.
 \label{phiaGeneral}
\end{eqnarray}
We consider here the case that there are two horizons, the event
horizon $r_+$ and Cauchy horizon $r_{-}$ (which are the solutions
of $F(r)=0$) for general spherically symmetric black holes. For
the nonextremal black hole, the Killing vector field $\phi^a$
(\ref{phiaGeneral}) has four zeros ($r_i,\theta_i$) located at
\begin{equation}
  (r_{+},0), ~~~ (r_{+},\pi), ~~~
  (r_{-},0), ~~~ (r_{-},\pi).
\label{zerosRN}
\end{equation}
For these zeros, the corresponding Hopf indexes and the Brouwer
degrees ($\beta_i, \eta_i$) are
\begin{equation}
 (1,+1),~~(1,+1),~~(1,-1),~~(1,-1).
\end{equation}
Again, it is only the zeros ($r_{+},0$) and ($r_{+},\pi$) that
have contribution to $\chi$, $\sigma$ and $\rho$. And the Euler
characteristic, the absorption cross section and its density are
the same as the case of the nonextremal RN black hole.
\begin{eqnarray}
 \chi =2,~
 \sigma = 2\frac{\cal A}{g_h},~
 \rho =\frac{\cal A}{g_h}~\delta(r-r_{+})
        \left[\delta(\theta)
        +\delta(\theta-\pi)\right]. \label{ourChiSpherically1}
\end{eqnarray}

For extremal black holes ($r_{+}=r_{-}$), the Killing vector field
$\phi^a$ has two zeros located at $(r_{+},0)$ and $(r_{+},\pi)$,
and the corresponding Hopf indices are $ \beta_1=\beta_2=0$, which
gives the following result
\begin{equation}
 \chi=0,~~~ \sigma=0,~~~ \rho=0. \label{ourChiSpherically2}
\end{equation}
This result is in agreement with the statements that the extremal
black hole has a unique internal state \cite{Hawking1995} and its
temperature is zero due to the vanishing of surface gravity on
horizon.

\subsection{The Kerr black hole}
In this subsection, we extend this work to the Kerr black hole.
The Kerr solution describes both the stationary axisymmetric
asymptotically flat gravitational field outside a massive rotating
body and a rotating black hole with mass $M$ and angular momentum
$J$. The Kerr black hole can also be viewed as the final state of
a collapsing star, uniquely determined by its mass and rate of
rotation. Moreover, its thermodynamical behavior is very different
from the Schwarzschild black hole or the RN black hole, because of
its much more complicated causal structure. Hence its study is of
great interest in understanding physical properties of
astrophysical objects, as well as in checking any conjecture about
thermodynamical properties of black holes.

In terms of Boyer--Lindquist coordinates, the Euclidean Kerr
metric reads \cite{Oneill}
\begin{eqnarray}
 ds^{2} &=& {\Delta \over \varrho^{2}}
           \left(dt-a{\sin^{2}\!\theta}d\varphi\right)^{2}
           + {\varrho^{2} \over \Delta}dr^{2}\cr
      && +\varrho^{2}d\theta^{2}
          +{\sin^{2}\theta \over \varrho^{2}}
          \Bigr[\left( r^{2}+a^{2}\right)d\varphi -adt\Bigr]^{2},
          \label{kermet}
\end{eqnarray}
where $\Delta$ is the Kerr horizon function
\begin{eqnarray}
\Delta = r^{2}-2Mr +a^{2},
\end{eqnarray}
and
\begin{eqnarray}
\varrho = r^{2} + a^{2}\cos^{2}\theta.
\end{eqnarray}
Here $a$ is the angular momentum for unit mass as measured from
the infinity; it vanishes in the Schwarzschild limit. The
nonextremal Kerr black hole has the event horizon $r_{+}$ and the
Cauchy horizon $r_{-}$ at
\begin{equation}
r_{\pm}=M \pm \sqrt{M^2-a^2},
\end{equation}
The extreme case corresponds to $r_{+}=r_{-}$ or $M=a$. Using the
GBC theorem, it has been proved
\cite{WangBinPRD1998,WangBinPLB1998} that $\chi=2$ for the
nonextremal Kerr black hole and $\chi=0$ for the extremal one.

Such a metric (\ref{kermet}) corresponds to the following vierbein
1-forms:
\begin{eqnarray}
 e^{0}&=&{{\sqrt\Delta} \over \varrho}
        \left(dt-a\sin^{2}\!\theta d\varphi\right), ~~~ 
 e^{1} = {\varrho \over{\sqrt\Delta}}dr, \label{vierb1} \\
 e^{2}&=&\varrho d\theta, ~~~
 e^{3}= {\sin\theta\over \varrho}
        \Bigr[\left(r^{2}+a^{2}\right)d\varphi
        -adt\Bigr],\label{vierb2}
\end{eqnarray}
where $\varrho$ is the positive square root of $\varrho^2$. The
Killing equation and the condition for the Killing vector field
$\phi^\mu$ are same as Eqs. (\ref{KillingEq}) and
(\ref{KillingEqCondiction}). By the relation
$\phi^a=e^a_{\mu}\phi^{\mu}$ and the vierbein given in
(\ref{vierb1}) and (\ref{vierb2}), the solution for $\phi^a$ is
read as
\begin{eqnarray}
 \phi^0=\sqrt{\frac{\Delta}{\varrho}},~~~
 \phi^1=\phi^2=0,~~~
 \phi^3=\frac{a\sin\theta}{\sqrt{\varrho}}.
 \label{phiaKerr}
\end{eqnarray}

For the nonextremal Kerr black hole, the Killing vector field
$\phi^a$ (\ref{phiaKerr}) has four zeros ($r_i,\theta_i$) located
at
\begin{equation}
  (r_{+},0), ~~~ (r_{+},\pi), ~~~
  (r_{-},0), ~~~ (r_{-},\pi).
\label{zerosRN}
\end{equation}
The Hopf indexes and the Brouwer degrees ($\beta_i, \eta_i$)
correspond to these zeros are
\begin{equation}
 (1,+1),~~(1,+1),~~(1,-1),~~(1,-1).
\end{equation}
Again, it is only the zeros ($r_{+},0$) and ($r_{+},\pi$) that
have contribution to the Euler characteristic $\chi$ which is
calculated to be
\begin{equation}
 \chi =2. \label{chiKerr1}
\end{equation}

For the extremal Kerr black hole, the Killing vector field
$\phi^a$ has two zeros located at $(r_{+},0)$ and $(r_{+},\pi)$,
and the corresponding Hopf indices are $ \beta_1=\beta_2=0$, which
gives the Euler characteristic
\begin{equation}
 \chi =0.  \label{chiKerr2}
\end{equation}


Lastly, let us give a comment on a more general spacetime--the
Kerr--Newman black hole, which metric has Petrov type $D$ and has
also the form of the Kerr black hole (\ref{kermet}) but with
$\Delta = r^{2}-2Mr +a^{2}+Q^2$. From the formalized solution of
the Killing vector field for the Kerr black hole (\ref{phiaKerr}),
one can conclude that the results (\ref{chiKerr1}) and
(\ref{chiKerr2}) are also held for the Kerr--Newman black hole. we
will assume that the result (\ref{blfermion}) generalizes to
arbitrary charged rotating black holes and so the relations
(\ref{ourChiSpherically1}) and (\ref{ourChiSpherically2})
continues to hold for the Kerr black hole and the Kerr--Newman
black hole.


\section{Conclusions}\label{SecConclusions}
In summary, we first present a relation between the low-energy
absorption cross section for minimally coupled fermions and the
Euler characteristic of (3+1)-dimensional spherically symmetric
black holes. From the relation, one can see clearly that the
topologies (\ref{nonextremal}) and (\ref{extremal}) of spherically
symmetric black holes, which correspond to the nonextremal and
extremal black holes, respectively, lead directly to the nonzero
and zero results of the absorption cross section. Then using the
relation and the GBC theorem, we introduce the absorption cross
section density to describe the relation between them. It is shown
that the absorption cross section and its density are determined
by the singularities of the timelike Killing vector field of
spacetime and these singularities are labeled by the topological
numbers, Hopf indices and Brouwer degrees. For nonextremal black
holes, there are two singularities with the Hopf indices $\beta=1$
and Brouwer degrees $\eta=+1$ on the north pole and south pole of
event horizon and two singularities with $\beta=1$  and $\eta=-1$
on the north pole and south pole of Cauchy horizon. In this case,
only the singularities on event horizon have contribution to the
Euler characteristic and the absorption cross section and one
obtains $\chi=2$ and $\sigma=2 g_h^{-1}{\cal A}$. For extremal
ones, there are only two singularities with $\beta=0$ on the north
pole and south pole of horizon. In this case, the singularities
have no contribution to the Euler characteristic and the
absorption cross section and then $\chi=0$ and $\sigma=0$. The
results (\ref{density_beta_eta}) and (\ref{CrossSection_beta_eta})
give information of quantization through the singularities and the
topological numbers, which can be looked upon as the topological
quantization of the absorption cross section.

\section*{Acknowledgment}
It is a pleasure to thank Prof. Guohong Yang for interesting
discussions. This work was supported by the National Natural
Science Foundation of the People's Republic of China and the
Fundamental Research Fund for Physics and Mathematic of Lanzhou
University.

\end{document}